\documentclass[journal]{IEEEtran}
\IEEEoverridecommandlockouts                              

\usepackage{multicol}
\usepackage{graphicx}
\usepackage{etoolbox}
\usepackage{amsmath} 
\usepackage{cite}
\usepackage[table,x11names]{xcolor}
\usepackage{soul,color}
\soulregister\cite7
\soulregister\ref7
\soulregister\pageref7
\usepackage{subcaption}
\usepackage[font=footnotesize]{caption}
\usepackage{epstopdf}
\DeclareGraphicsExtensions{.eps}

\begin{document}
\title{Multiscale Fluctuation-based Dispersion Entropy and its Applications to Neurological Diseases}

\author{Hamed Azami$^{1,*}$, Steven E. Arnold$^{2}$, Saeid Sanei$^{3}$, \textit{Senior Member},\textit{ IEEE}, Zhuoqing Chang$^{4}$, Guillermo Sapiro$^{5}$, \textit{Fellow}, \textit{IEEE}, Javier Escudero$^{6}$, \textit{Member},\textit{ IEEE}, and Anoopum S. Gupta$^{7}$
\thanks{$^{1}$Department of Neurology and Massachusetts General Hospital, Harvard University, Charlestown, MA 02129, USA.
	
*Corresponding author, email: hazami@mgh.harvard.edu.}
\thanks{$^{2}$S. E. Arnold is with the Department of Neurology and Massachusetts General Hospital, Harvard Medical School, Charlestown, MA 02129, USA.}
\thanks{$^{3}$S. Sanei is with the School of Science and Technology, Nottingham Trent University, Nottingham, UK.}
\thanks{$^{4}$Z. Chang is with the Department of Electrical and Computer Engineering, Duke University, Durham, NC 27707, USA.}
\thanks{$^{5}$G. Sapiro is with the Department of Electrical and Computer Engineering, Computer Sciences, Biomedical Engineering, and Math, Duke University, Durham, NC 27707, USA.}
\thanks{$^{6}$J. Escudero is with the Institute for Digital Communications, School of Engineering, University of Edinburgh, Edinburgh, UK.}
\thanks{$^{7}$A. S. Gupta is with the Department of Neurology and Massachusetts General Hospital, Harvard Medical School, Boston, MA 02114, USA.}}

\maketitle

\begin{abstract}
Fluctuation-based dispersion entropy (FDispEn) is a new approach to estimate the dynamical variability of the fluctuations of signals. It is based on Shannon entropy and fluctuation-based dispersion patterns. To quantify the physiological dynamics over multiple time scales, multiscale FDispEn (MFDE) is developed in this article. MFDE is robust to the presence of baseline wanders, or trends, in the data. We evaluate MFDE, compared with popular multiscale sample entropy (MSE), and the recently introduced multiscale dispersion entropy (MDE), on selected synthetic data and five neurological diseases' datasets: 1) focal and non-focal electroencephalograms (EEGs); 2) walking stride interval signals for young, elderly, and Parkinson's subjects; 3) stride interval fluctuations for Huntington's disease and amyotrophic lateral sclerosis; 4) EEGs for controls and Alzheimer's disease patients; and 5) eye movement data for Parkinson's disease and ataxia. MFDE dealt with the problem of undefined MSE values and, compared with MDE, led to more stable entropy values over the scale factors for pink noise. Overall, MFDE was the fastest and most consistent method for the discrimination of different states of neurological data, especially where the mean value of a time series considerably changes along the signal (e.g., eye movement data). This study shows that MFDE is a relevant new metric to gain further insights into the dynamics of neurological diseases recordings.
\end{abstract}


\maketitle

\section{Introduction}
One of the most popular and powerful nonlinear measures used to evaluate the dynamical characteristics of signals is entropy \cite{pincus1991approximate,richman2000physiological,bandt2002permutation,rostaghi2016dispersion}. Shannon entropy (ShEn) and conditional entropy (ConEn) are two key fundamental concepts in information theory widely used for characterization of physiological signals \cite{richman2000physiological,bandt2002permutation}. ShEn and ConEn show the amount of information and rate of information production, respectively, and are related to the uncertainty or irregularity of data \cite{richman2000physiological,bandt2002permutation,rostaghi2016dispersion,shannon2001mathematical}. A higher entropy value demonstrates higher irregularity, while smaller entropy values show lower irregularity or uncertainty in a time series \cite{richman2000physiological,rostaghi2016dispersion,sanei2007eeg}.

Existing entropy techniques, such as sample entropy (SampEn) and permutation entropy (PerEn), are widely used to quantify the irregularity of signals at one temporal scale \cite{shannon2001mathematical,rostaghi2016dispersion}. They assess repetitive patterns and return maximum values for completely random processes \cite{fogedby1992phase,costa2005multiscale,zhang1991complexity}. However, these techniques fail to account for the multiple time scales inherent in biomedical recordings \cite{costa2005multiscale,azami2017refined_MFE}. To deal with this limitation, multiscale SampEn (MSE) was proposed \cite{costa2002multiscale} and it has become a prevalent algorithm to quantify the complexity of univariate time series, especially physiological recordings \cite{humeau2015multiscale,costa2005multiscale}.

Following \cite{costa2002multiscale,costa2005multiscale}, the concept of complexity stands for ''meaningful structural richness'', which may be in contrast with uncertainty or irregularity of time series defined by classical entropy approaches such as SampEn and PerEn \cite{bar1997dynamics,costa2005multiscale,fogedby1992phase,rostaghi2016dispersion}. As mentioned above, these entropy approaches evaluate repetitive patterns and return maximum values for completely random processes \cite{fogedby1992phase,costa2005multiscale,zhang1991complexity}. However, a completely ordered time series with a low entropy value or a completely disordered signal with a high entropy value is the least complex \cite{fogedby1992phase,costa2005multiscale,silva2015multiscale}. For instance, white noise is more irregular than pink noise ($1/f$ noise) even though the latter is more complex since the pink noise has long-range correlations and its $1/f$ decay produces a fractal structure in time \cite{fogedby1992phase,costa2005multiscale,silva2015multiscale}.  

In brief, the concept of complexity builds on three hypotheses: I) the complexity of a physiological time series indicates its ability to adapt and function in ever-changing environment; II) a biological time series requires to operate across multiple temporal scales and so, its complexity is similarly multiscaled and hierarchical; and III) a wide class of disease states, in addition to aging, decrease the adaptive capacity of the individual, thus reducing the information carried by output variables. Therefore, the MSE focuses on quantifying the information expressed by the physiologic dynamics over multiple temporal scales \cite{costa2005multiscale,fogedby1992phase}.

In spite of its popularity, MSE is undefined or unreliable for very short signals and computationally complex for real-time applications as a result of using SampEn \cite{azami2017refined_MFE,morabito2012multivariate}. To address these shortcomings, multiscale PerEn (MPE) was proposed \cite{morabito2012multivariate}. Although MPE is able to deal with short signals and is considerably faster than MSE, it does not fulfill the key hypotheses of the concept of complexity as described above \cite{wu2016refined}. Furthermore, the behaviour of MPE is different from that of MSE in some cases so, in reality, it is not a replacement. To overcome the limitations of MPE and MSE at the same time, we have recently introduced multiscale dispersion entropy (DispEn - MDE), based on our developed DispEn \cite{rostaghi2016dispersion,azami2018amplitude}, to quantify the complexity of signals \cite{azami2017refined}.

Compared with the conventional complexity approaches, 1) MDE increases the reliability of the results and at the same time does not lead to undefined values for short signals, 2) MDE is markedly faster, especially for long signals, and 3) it yields larger differences between physiological conditions, such as subjects with epilepsy disorders or Alzheimer's disease (AD) vs.~matched controls \cite{azami2017refined}. 

MSE and MDE have been applied in different research fields, including biomedical engineering and neuroscience \cite{humeau2015multiscale,azami2018coarse}. MSE was successfully used for the diagnosis of depression using heart rate variability, speech recordings, and electroencephalograms (EEGs) \cite{goldberger2014complexity}. Using MSE, an increased EEG signal complexity was found in Parkinson's disease (PD) patients during non-rapid eye movement sleep at high scale factors \cite{chung2013multiscale}. MDE was successfully used for sleep stage classification using single-channel electrooculography signals \cite{rahman2018sleep}. Miskovic \textit{et al.} showed that slow sleep EEG data were characterized by reduced MDE values at low scales and increased MDE values at high scale factors \cite{miskovic2019changes}. MDE and MSE were used to discriminate AD patients from age-matched controls using magnetoencephalogram signals \cite{azami2017multiscale}. The differences between the MDE values for the AD vs.~healthy subjects were more significant than their corresponding MSE-based values.

In many real-world applications (e.g., in computing the correlation function and in spectral analysis), the (local or global) trends from a signal \cite{hu2001effect,wu2007trend} need to be removed. In such methods, after detrending the local or global trends of a time series, the fluctuations are evaluated \cite{hu2001effect,wu2007trend}. When only the fluctuations of data are relevant or local trends of a time series are irrelevant \cite{hu2001effect,wu2007trend,peng1995quantification}, there is no difference between dispersion patterns $\{11\}$ , $\{22\}$ , and $\{33\}$ or $\{12\}$ and $\{23\}$. That is, the fluctuations of $\{11\}$, $\{22\}$ , and $\{33\}$ or $\{12\}$ and $\{23\}$ are equal. Thus, we have very recently introduced fluctuation-based DispEn (FDispEn) \cite{azami2018amplitude}. The potential of FDispEn for characterization of various synthetic and biomedical data was shown. For example, FDispEn significantly discriminated eleven 3-4 years old children from twelve 11-14 years old subjects using their stride interval fluctuations \cite{azami2018amplitude}. However, this was never extended to multiscale for covering a wider range of applications.

Therefore, the main contributions of this study are proposing multiscale FDispEn (MFDE) and evaluating MFDE, MDE, and MSE on selected synthetic signals and five neurological datasets: focal and non-focal EEGs, stride interval fluctuations in PD, young and elderly individuals as well as Huntington's disease (HD) and amyotrophic lateral sclerosis (ALS), resting-state EEG activity in AD, and eye movement data in ataxia vs.~PD.
	
This article is structured as follows. In Section II, the MFDE algorithm is detailed. The synthetic and real datasets used here are briefly described in Section III. The results and discussion are provided in Section IV. After describing future works in Section V, we conclude the paper in Section VI.

\section{Methods}
\subsection{Multiscale Fluctuation-based Dispersion Entropy (MFDE)}
MFDE is based on the coarse-graining process \cite{costa2005multiscale} and FDispEn \cite{azami2018amplitude}. Assume we have a univariate signal of length $L$: $\textbf{u}=\{\textit{u}_{1}, \textit{u}_{2},..., \textit{u}_{L}\}$. In the MFDE algorithm, the original signal \textbf{u} is first divided into non-overlapping segments of length $\tau$, named scale factor. Afterwards, the average of each segment is calculated to derive a coarse-grained time series as follows \cite{costa2005multiscale}:
\begin{equation} 
{{x}_{j}}^{(\tau )}=\frac{1}{\tau }\sum\limits_{b=(j-1)\tau +1}^{j\tau }{{{u}_{b}}}, \quad 1\le j\le \left\lfloor \frac{L}{\tau } \right\rfloor =N\,
\end{equation}
Of note is that other coarse-graining processes can be used in this step \cite{azami2018coarse}, but, for the sake of clarity, we focus on the original definition in this paper. Finally, the FDispEn of each coarse-grained signal ${{x}_{j}}^{(\tau )}$ is calculated.

The FDispEn of the univariate signal of length $N$: $\textbf{x}=\{\textit{x}_{1}, \textit{x}_{2},..., \textit{x}_{N}\}$ is defined as follows:

Step 1) First, $x_{j} (j=1,2,...,N)$ are mapped to $\textit{c}$ classes with integer indices from 1 to $\textit{c}$. To this end, the normal cumulative distribution function (NCDF) is first utilized to overcome the problem of assigning the majority of $x_{i}$ to only few classes, especially when thr maximum or minimum values are noticeable larger or smaller than the mean/median value of the signal \cite{azami2018amplitude,rostaghi2016dispersion,azami2017refined}. For more information about the reasons behind using NCDF, please see \cite{azami2018amplitude,rostaghi2016dispersion}.

The NCDF maps $\textbf{x}$ into $\textbf{y}=\{\textit{y}_{1}, \textit{y}_{2},..., \textit{y}_{N}\}$ from 0 to 1 as follows: 
\begin{equation}
y_{j}=\frac{1}{\sigma\sqrt{2\pi}}\int\limits_{-\infty}^{x_{j}}{e}^{\frac{-(t-\mu)^2}{2\sigma^2}}\,\mathrm{d}t,
\end{equation}
where $\sigma$ and $\mu$ are the SD and mean of time series \textbf{x}, respectively. Then, we linearly assign each $y_{i}$ to an integer from 1 to $c$. To do so, for each member of the mapped signal, we use $z_{j}^{c}=\mbox{round}(c\cdot y_{j}+0.5)$, where  $z_{j}^{c}$ denotes the $j^{th}$ member of the classified time series and the rounding operator involves either increasing or decreasing a number to the next digit \cite{azami2018amplitude,rostaghi2016dispersion,azami2017refined}.

Step 2) Time series $\textbf{z}^{m,c}_i$ are defined with respect to embedding dimension $m-1$ and time delay $d$ according to $\textbf{z}^{m,c}_i=\{z^{c}_i,{z}^{c}_{i+d},...,{z}^{c}_{i+(m-1)d}\} $, $i=1,2,...,N-(m-1)d$ \cite{rostaghi2016dispersion,azami2018amplitude}. Each time series $\textbf{z}^{m,c}_i$ is mapped to a fluctuation-based dispersion pattern ${{\pi }_{{{v}_{0}}{{v}_{1}}...{{v}_{m-1}}}}$, where $z_{i}^{c}=v_0$, $z_{i+d}^{c}=v_1$,..., $z_{i+(m-1)d}^{c}=v_{m-1}$. The number of possible fluctuation-based dispersion patterns that can be assigned to each time series $\textbf{z}^{m,c}_i$ is equal to $(2c-1)^{(m-1)}$ \cite{azami2018amplitude}. 

Step 3) For each $(2c-1)^{m-1}$ potential dispersion patterns ${{\pi }_{{{v}_{0}}...{{v}_{m-1}}}}$, relative frequency is obtained as follows:
\begin{equation}
\begin{split}
&p({{\pi }_{{{v}_{0}}...{{v}_{m-1}}}})=\\
&\frac{\#\{i\left| i\le N-(m-1)d,\mathbf{z}_{i}^{m,c}\text{ has type }{{\pi }_{{{v}_{0}}...{{v}_{m-1}}}} \right.\}}{N-(m-1)d},
\end{split}
\end{equation}

\setlength{\parindent}{0pt}%
where $\#$ means cardinality. In fact, $p({{\pi }_{{{v}_{0}}...{{v}_{m-1}}}})$ shows the number of dispersion patterns of ${{\pi }_{{{v}_{0}}...{{v}_{m-1}}}}$ that is assigned to $\textbf{z}^{m,c}_i$, divided by the total number of embedded signals with embedding dimension $m$.
\setlength{\parindent}{9pt}

Step 4) Finally, based on Shannon's definition of entropy, the FDispEn value is calculated as follows:
\begin{equation}\label{eq_2}
\begin{split}
&FDispEn(\mathbf{x},m,c,d)=\\
&-\sum_{\pi=1}^{(2c-1)^{m-1}} {p({{\pi}_{{{v}_{0}}...{{v}_{m-1}}}})\cdot\ln }\left(p({{\pi}_{{{v}_{0}}...{{v}_{m-1}}}}) \right),   
\end{split}
\end{equation}

It is worth noting that the mapping based on the NCDF used in the calculation of FDispEn \cite{rostaghi2016dispersion} for the first temporal scale is maintained across all scales. In fact, in MFDE, $\mu$ and $\sigma$ of NCDF are respectively set at the average and standard deviation (SD) of the original signal and they remain constant for all scale factors. This approach is similar to keeping $r$ constant (usually 0.15 of the SD of the original signal) in the MSE-based algorithms \cite{costa2005multiscale}.

\begin{figure*}
	\centering
	\includegraphics[width=17cm]{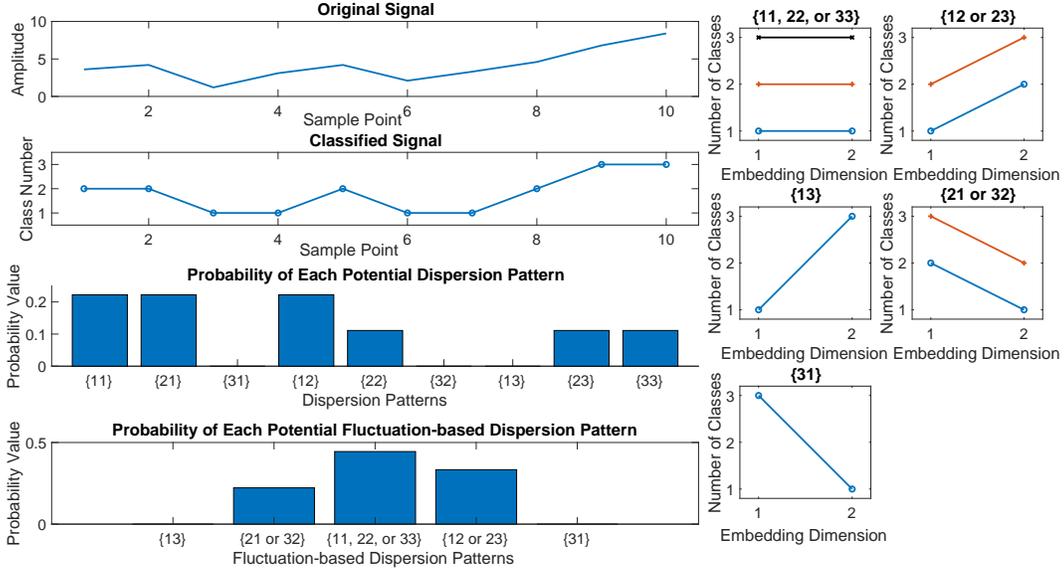}
	\caption{Illustration of the FDispEn vs.~DispEn algorithms using linear mapping of $\textbf{x}=\{3.6,4.2,1.2,3.1,4.2,2.1,3.3,4.6,6.8,8.4\}$ (top left) with the time delay 1, number of classes 3, and embedding dimension 2. The nine dispersion patterns $\{11,12,13,21,22,23,31,32,33\}$ and five fluctuation-based dispersion patterns $\{11,12,13,21,31\}$ are shown on the right of Figure. The relative frequency for both DispEn and FDispEn are illustrated on the bottom left of Figure.}
	\label{Plot_MFDE_MDE}
\end{figure*}

FDispEn deals with the differences between adjacent elements of dispersion patterns, named fluctuation-based dispersion patterns \cite{azami2018amplitude}. In this way, we have vectors with length $m-1,$ which each of their elements changes from $-c+1$ to $c-1$. Thus, there are $(2c-1)^{m-1}$ potential fluctuation-based dispersion patterns. For instance, let us have a series $\textbf{x}=\{3.6, 4.2, 1.2, 3.1, 4.2, 2.1, 3.3, 4.6, 6.8, 8.4\}$, shown on the top left of Fig.~\ref{Plot_MFDE_MDE}. We want to calculate the FDispEn value of \textbf{x}. For simplicity, we set $d=1$, $m=2$, and $c=3$. The five potential fluctuation-based dispersion patterns vs.~nine potential dispersion patterns are depicted on the right of Fig.~\ref{Plot_MFDE_MDE}. Step 1: $x_j$ ($j=1,2,\dots,10$) are linearly mapped into three classes with integer indices from 1 to 3, as can be seen in Fig.~\ref{Plot_MFDE_MDE}. Step 2: a window with length 2 (embedding dimension) moves along the signal and the number of each of the fluctuation-based dispersion patterns is counted. Step 3: the relative frequency for both DispEn and FDispEn are shown on the bottom left of Fig.~\ref{Plot_MFDE_MDE}. Step 4: using Equation (\ref{eq_2}), the FDispEn value of \textbf{x} is equal to $-(\frac{4}{9}\ln(\frac{4}{9})+\frac{3}{9}\ln(\frac{3}{9})+\frac{2}{9}\ln(\frac{2}{9}))=1.0609$.

When all possible fluctuation-based dispersion patterns have equal probability value, the highest value of FDispEn is obtained, which has a value of $ln ({(2c-1)^{m-1}})$. In contrast, if there is only one $ p({{\pi }_{{{v}_{0}}...{{v}_{m-1}}}}) $ different from zero, which demonstrates a completely regular/predictable time series, the smallest value of FDispEn is obtained \cite{azami2018amplitude}.

\subsection{Parameters of MFDE}\label{Parameters}
There are four parameters for MFDE, namely the embedding dimension \textit{m}, the number of classes \textit{c}, the time delay \textit{d}, and the maximum scale factor $\tau_{max}$.

Based on the existing complexity-based approaches \cite{costa2002multiscale,costa2005multiscale,morabito2012multivariate,azami2017refined_MFE}, the time delay was set equal to 1 in this study. However, if the sampling frequency is noticeably larger than the highest frequency component of a signal, the first minimum or zero crossing of the autocorrelation function or mutual information can be used for the selection of an appropriate time delay \cite{kaffashi2008effect,azami2018coarse}.

It is considered that $c>1$ to avoid the trivial case of having only one fluctuation-based dispersion pattern. For MFDE and MDE, we set $c=6$ for all the data in this work, albeit the range $2<c<9$ leads to similar findings. For more information about \textit{c}, \textit{m}, and \textit{d}, please refer to \cite{azami2018amplitude,azami2017refined}. 

To work with reliable statistics to calculate FDispEn, it is recommended that the number of potential fluctuation-based dispersion patterns is smaller than the length of the signal ($ (2c-1)^{m-1}<L$) \cite{azami2018amplitude}. For MFDE, the coarse-graining process causes the length of a signal decreases to $\left\lfloor \frac{L}{\tau_{max} } \right\rfloor$, it is recommended to have $(2c-1)^{m-1}<\left\lfloor \frac{L}{\tau_{max} } \right\rfloor$. 

For all the following experiments, we set $m=2$ and $d=1$ for MFDE, MDE, and MSE. The number of classes is equal to 6 for both the MDE and MFDE techniques. The threshold \textit{r} for MSE, which is used as a benchmark, was chosen as 0.15 of the SD of a signal \cite{costa2005multiscale}. Finally, for consistency, the maximum scale factor $\tau_{max}$ was set based on $c^{m}<\left\lfloor \frac{L}{\tau_{max} } \right\rfloor$ for all the complexity techniques used herein \cite{azami2017refined}.

\section{Evaluation Signals}
To assess the ability of MFDE, compare it with MSE and MDE, and to characterize various univariate time series, we use the following synthetic and neurological datasets.

\subsection{Synthetic Signals} \label{SynthSiG}
1) The complexity of pink noise ($1/f$ noise) is higher than white noise, whereas the irregularity or uncertainty of the former signal is lower than the latter \cite{costa2005multiscale,fogedby1992phase,azami2017refined}. Thus, white and pink noise are two suitable data for assessing the multiscale entropy techniques \cite{costa2005multiscale,fogedby1992phase,silva2015multiscale,wu2016refined,humeau2016refined}. For more information about white vs.~pink noise, please refer to \cite{costa2005multiscale,azami2016improved}.

2) Physiological signals are often corrupted by different kinds of noise, such as additive white Gaussian noise (WGN) \cite{lam2011preserving}. A WGN is also considered as a basic statistical model used in information theory to mimic the effect of random processes that occur in nature \cite{houdrehigh2016}. In order to understand the relationship between MFDE, MSE, and MDE, and the level of noise affecting periodic time series, we generated an amplitude-modulated periodic signal with a WGN with diverse power. First, we generated a time series as an amplitude-modulated sum of two cosine waves with frequencies at 0.5 Hz and 1 Hz. The first 20 s of this series (100 s) does not have any noise. Then, WGN was added to the time series \cite{azami2016improved}.

\subsection{Neurological Datasets}
Discrimination of people with neurological diseases from healthy subjects, or among different neurological diseases, by analysis of their recorded time series is a long-standing challenge in the physiological complexity literature \cite{costa2005multiscale,labate2013entropic,yang2013cognitive,andrzejak2012nonrandomness,azami2017refined,chung2013multiscale}. EEGs, walking stride interval time series, and eye movement are clinical pavements that may be helpful in diagnosis and tracking of neurological diseases states \cite{sanei2007eeg,hausdorff1996fractal,andrzejak2012nonrandomness,azami2017refined}. Using these recordings, MFDE, MDE, and MSE are used to characterize several neurological diseases such as ALS, AD, PD, cerebellar ataxias, and HD.

\textit{1) Dataset of Focal and Non-focal Electroencephalograms (EEGs)}: Epilepsy is a common neurological condition. EEG signals are used to identify areas that generate or propagate by seizures \cite{andrzejak2012nonrandomness,acharya2018characterization}. Generally, focal EEG signals are recorded from the epileptic part of the brain, whereas non-focal EEGs correspond to brain regions unaffected by epilepsy \cite{acharya2018characterization}. The ability of MFDE, MDE, and MSE to discriminate focal from non-focal signals is evaluated by the use of an EEG dataset (publicly-available at \cite{Ralph_Data}) \cite{andrzejak2012nonrandomness}.

\begin{figure}
	\centering
	\includegraphics[width=8cm]{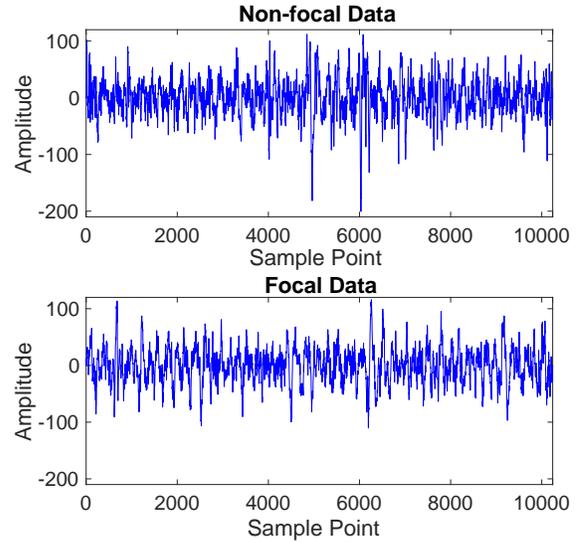}
	\caption{Example of a focal and non-focal EEG time series. }
	\label{Signal_NonFocal}
\end{figure}

The dataset includes 5 patients and, for each patient, there are 750 focal and 750 non-focal bivariate time series. The length of each signal was 20 s with sampling frequency of 512 Hz (10240 samples). Focal and non-focal EEG time series samples are depicted in Fig.~\ref{Signal_NonFocal}. For more information, please, refer to \cite{andrzejak2012nonrandomness}. All subjects gave written informed consent that their signals from long-term EEG might be used for research purposes \cite{andrzejak2012nonrandomness}. Before applying the complexity methods, the time series were digitally filtered using a Hamming window FIR band-pass filter of order 200 and cut-off frequencies 0.5 Hz and 40 Hz, a band typically used in the analysis of brain activity. 

\textit{2) Dataset of Walking Stride Interval Time Series for Young, Elderly, and Parkinson's Disease (PD) Subjects}: 
It was shown that aging leads to less complex recordings of stride \cite{costa2005multiscale,hausdorff1996fractal}. It was also documented that the gait of ALS patients is less stable and more temporally disorganized in comparison with that of healthy individuals. Furthermore, advanced ALS, HD, and PD were associated with certain common, but also distinct, features of altered stride dynamics \cite{hausdorff1996fractal,hausdorff2000dynamic}. To this end, we use the walking stride interval fluctuations to distinguish PD patients from healthy elderly subjects, young from elderly people, and ALS from HD patients (next dataset).

\begin{figure}
	\centering
	\includegraphics[width=8cm]{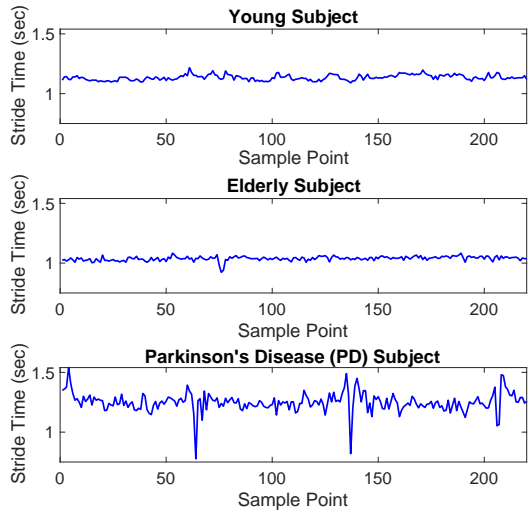}
	\caption{Example of effects of aging and Parkinson's disease on fluctuations of stride-interval dynamics.}
	\label{Strdie_Interval_PD_Young_Old_Signal}
\end{figure}

To compare MFDE, MDE, and MSE, publicly-available stride interval recordings were used \cite{hausdorff1996fractal,XYZ}. The signals were recorded from five young, healthy men (23 - 29 years old), five healthy old adults (71 - 77 years old), and five elderly adults (60 - 77 years old) with PD. All the individuals walked continuously on level ground around an obstacle-free path for 15 minutes. The stride interval was measured by the use of ultra-thin, force sensitive resistors placed inside the shoe. Fig.~\ref{Strdie_Interval_PD_Young_Old_Signal} shows an example of the stride-interval time series for a young, an elderly, and a PD subject. For more information, please refer to \cite{XYZ}.

\textit{3) Dataset of Walking Stride Interval Time Series for Huntington's Disease (HD) vs.~Amyotrophic Lateral Sclerosis (ALS) Patients}: 
 For the HD subjects, there is an increased randomness in stride interval fluctuations as compared with the healthy people \cite{hausdorff1996fractal,hausdorff2000dynamic}. On the other hand, gait usually becomes abnormal during the course of the ALS disease. A decreased (average) walking velocity was reported in ALS \cite{goldfarb1984gait}. It is yet unknown if the loss of motoneurons also changes the stride-to-stride complexity of gait. 

\begin{figure}
	\centering
	\includegraphics[width=8cm]{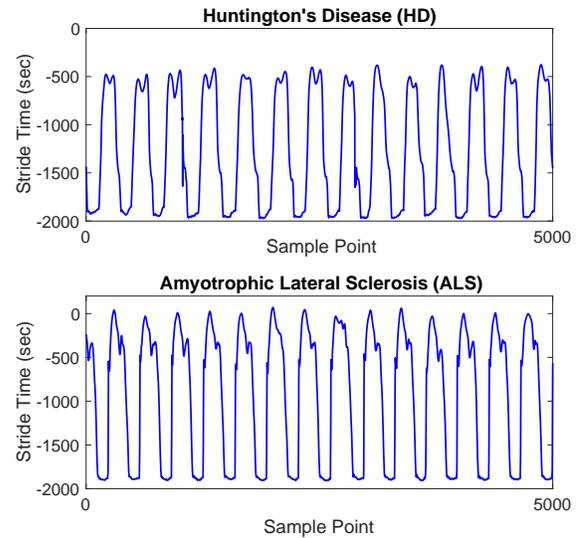}
	\caption{Example of effects of amyotrophic lateral sclerosis and Huntington's disease on fluctuations of stride-interval dynamics.}
	\label{Signal_HD_ALS}
\end{figure}

The records, which are available at \cite{ALS_HD}, are from 20 HD and 13 ALS patients. The mean age of the HD and ALS patients respectively were 47 (range 29-71) and 54.9 years (range 36-70). Subjects with ALS were able to walk independently for five minutes and did not use a wheelchair or assistive device for mobility. The subjects were instructed to walk at their normal pace along a 77-m-long hallway for 5 minutes. To measure the gait rhythm and the timing of the gait cycle, force-sensitive insoles were placed in the patients' shoes. The sampling frequency of the data was 300 Hz. Fig.~\ref{Signal_HD_ALS} shows an example of the stride-interval time series for a HD and an ALS subject. Note that all the patients provided informed, written consent and the study was approved by the Massachusetts General Hospital (MGH) Institutional Review Board. For more information about the dataset, please refer to \cite{hausdorff2000dynamic}.

\textit{4) Surface Electroencephalogram (EEG) Dataset of Brain Activity in Alzheimer's Disease (AD)}: 
AD, as a neurodegenerative disease, is the most common form of dementia \cite{alzheimer20172017,wilson2012natural}. AD changes the interaction between neurons in the brain during its progression. Consequently, it alters brain activity. Some of these changes may be recorded by the EEG technique \cite{dauwels2010diagnosis,bhat2015clinical,abasolo2006entropy,abasolo2008approximate}.

The 16-channel EEG dataset includes 11 AD patients (5 men; 6 women; age: 72.5 $\pm$ 8.3 years, all data given as mean $\pm$ SD) and 11 age-matched control healthy subjects (7 men; 4 women; age: 72.8 $\pm$ 6.1 years) \cite{escudero2006analysis}. To screen their cognitive status, a mini-mental state examination (MMSE) \cite{tombaugh1992mini} was done. The MMSE scores for AD patients and healthy subjects are 13.3 $\pm$ 5.6 and 30 $\pm$ 0, respectively. 

\begin{figure}
	\centering
	\includegraphics[width=8cm]{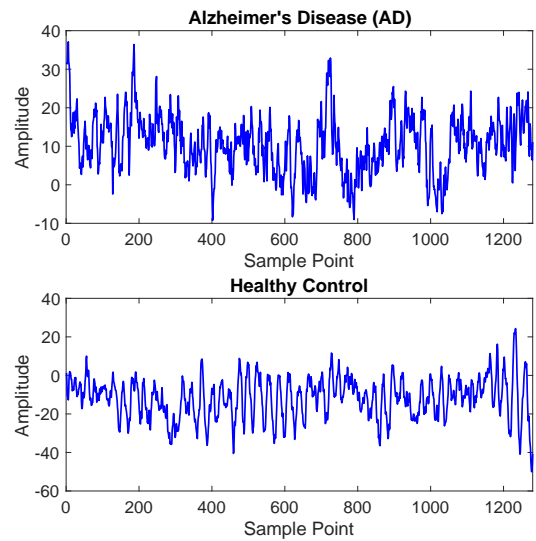}
	\caption{Example of effects of Alzheimer's disease on EEG time series.}
	\label{Signal_AD_Control}
\end{figure}

The subjects were recruited from the Alzheimer's Patients' Relatives Association of Valladolid (AFAVA), Spain. The	EEG time series were recorded with Oxford Instruments	Profile	Study Room 2.3.411 EEG equipment at	the	Hospital Cl\'inico Universitario de Valladolid (Spain). The EEGs were recorded using the international 10-20 system, in an eyes closed and resting state. All 16 electrodes were referenced to the linked ear lobes of	each individual. The signals were sampled at 256Hz and digitized with a 12-bit analog-to-digital converter. Informed consent was obtained for all 22 subjects and the local ethics committee approved the study. Before band-pass filtering with cut-off frequencies 1 and 40 Hz and a Hamming window with order 200, the signals were visually examined by an expert physician to select 5 s epochs (1280 samples) with minimal artifacts for analysis. On average, $30.0\pm12.5$ epochs (mean$\pm$SD) were selected from each electrode and each subject. An example of an AD EEG signal vs.~an age-matched healthy control's EEG is shown in Fig.~\ref{Signal_AD_Control}.

\textit{5) Eye Movement Dataset for Parkinsonism and Ataxia Patients}: 
Neurodegenerative diseases affect oculomotor function in a variety of ways, which impact vision and also provide clues into the underlying pathology and diagnosis. Cerebellar ataxias are an heterogeneous group of inherited and acquired diseases. As a broad group, ataxias cause profound and characteristic abnormalities in smooth pursuit, saccades, and fixation \cite{buttner98}. Oculomotor abnormalities in PD are clinically more subtle, but quantitative testing demonstrates abnormalities in both saccades and in smooth pursuit \cite{vidailhet94,rottach96}. 

\begin{figure}
	\centering
	\includegraphics[width=9cm]{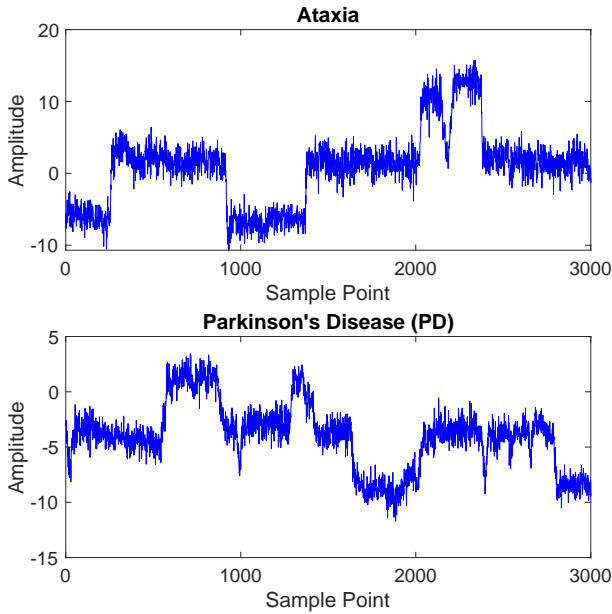}
	\caption{Example of eye movements for Parkinson's disease vs.~ataxia.}
	\label{Signal_Ataxia_PD}
\end{figure}

Participants with cerebellar ataxia and parkinsonism were recruited to participate in eye movement testing in MGH Neurology clinics. Stimuli for the antisaccades task were presented on an Apple iPad screen, while simultaneously recording each participant's face from an Apple iPhone camera sampling at 240fps. The video was processed using \cite{VIDEO} to extract facial landmarks, in particular the iris center. 57 participants with cerebellar ataxia and 20 participants with parkinsonism (18 with Parkinson's disease and 2 with atypical parkinsonism) were included in this dataset. An example of eye movements for Parkinson's disease vs.~ataxia is depicted in Fig.~\ref{Signal_Ataxia_PD}.

\section{Results and Discussion}
\subsection{Synthetic Signals} 
\begin{figure*}
	\centering
	\includegraphics[width=18cm]{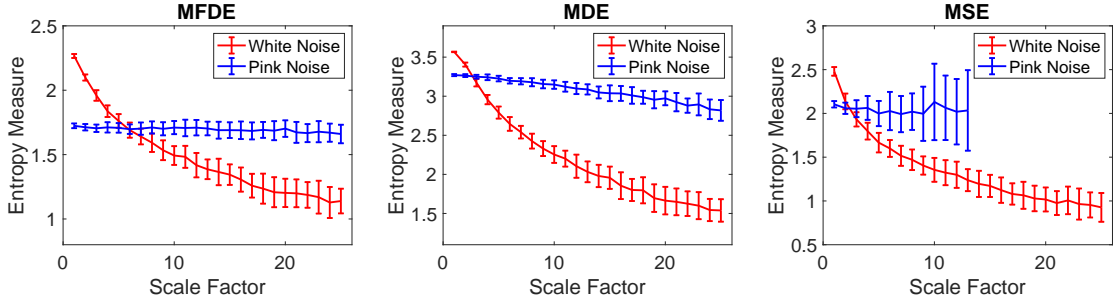}
	\caption{Mean value and SD of the MFDE, MDE, and MSE results for 40 different pink and white noise time series. The MSE values are undefined at several high scale factors.}
	\label{White_Pink_Noise}
\end{figure*}

Fig.~\ref{White_Pink_Noise} demonstrates the results obtained for MFDE, MDE, and MSE using 40 different white and pink noise signals with a length of 5,000 sample points. All the results are in agreement with the fact that pink noise has more complex structure than white noise, and white noise is more irregular than pink noise \cite{costa2005multiscale,silva2015multiscale,fogedby1992phase}. At short scale factors, the entropy values of white noise are higher than those of pink noise. At high scale factors the entropy value for the coarse-grained pink noise time series stays almost constant, whereas for the coarse-grained white noise data monotonically decreases. A slightly decreasing trend in MDE for pink noise is observed, but not so much in MFDE, showing an advantage of MFDE over MDE. For white noise, when the length of the signal, obtained by the coarse-graining process, decreases (i.e., the scale factor increases), the mean value of each segment converges to a constant value and the SD becomes smaller. Therefore, no new structures are revealed on higher scales. This demonstrates white noise signals contain information only at short time scales \cite{costa2005multiscale,silva2015multiscale}. For MSE, MDE and MFDE, we set $m=2$ and $d=1$, according to Subsection \ref{Parameters}.

\begin{figure*}
	\centering
	\includegraphics[width=18cm]{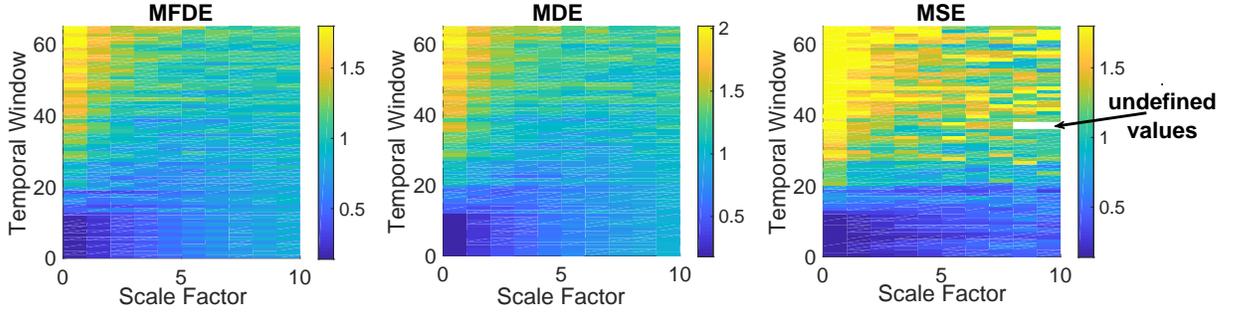}
	\caption{Mean value and SD of the MFDE, MDE, and MSE results for the quasi-periodic time series with increasing additive noise power using a window moving along the signal (temporal window). The MSE values at several temporal scale factors are undefined.}
	\label{Quasi_Periodic_AWGN}
\end{figure*}

The MFDE, MDE, and MSE methods are applied to the quasi-periodic signals with additive noise using a moving window of 450 samples (3 s) with 50\% overlap. Fig.~\ref{Quasi_Periodic_AWGN} demonstrates the MFDE-, MDE- and MSE-based profiles using the quasi-periodic signal with increasing additive noise power. As expected, the entropy values for all the three methods increase along the signal. At high scale factors, the entropy values decrease due to the filtering nature of the coarse-graining process \cite{azami2018coarse}. To sum up, the results show that all the methods lead to the similar findings, although the MDE and MFDE values are slightly more stable than the MSE ones, as demonstrated by the smoother nature of variations for MDE and MFDE, compared with MSE. Therefore, when a high level of noise is present, MDE and MFDE result in more stable profiles than MSE.  

To evaluate the computation time of MFDE (with \textit{m}=2 and 3 for completeness), MDE (\textit{m}=2 and 3), and MSE (\textit{m}=2 and 3), we use white noise signals with different lengths, changing from 100 to 100,000 sample points. The results are shown in Table I. The simulations were carried out using a PC with Intel (R) Xeon (R) CPU, E5420, 2.5 GHz and 8-GB RAM by MATLAB R2015a. For 100 and 300 sample points, MSE ($m=2$ and 3) results in undefined values at least at several scale factors. This does not happen for MDE and MFDE, demonstrating the advantage of these methods over MSE for short time series. There is no major difference between the computation time for the MSE with \textit{m}=2 and 3. The results show that for the different number of sample points, MFDE and MDE are considerably faster than MSE. This computational advantage of MFDE and MDE increases markedly with the data length. It is consistent with the fact that the computational cost of SampEn, FDispEn, and DispEn are O($N^{2}$), O($N$), and O($N$), respectively \cite{rostaghi2016dispersion,azami2018amplitude,wu2016refined}. Note that the MSE and MDE codes used in this paper are publicly-available at http://dx.doi.org/10.7488/ds/1477 and http://dx.doi.org/10.7488/ds/1982, respectively.

\begin{table*}
	\label{tab:table4}\caption{The computational time of MFDE, MDE, and MSE.} 
	\setlength{\tabcolsep}{3pt}
	\begin{tabular}{c*{8}{c}}
		Number of samples $\rightarrow$   &100& 300& 1,000& 3,000& 10,000& 30,000&100,000\\
		
		\hline
		MFDE($m=2$)          & 0.0028 s        &   0.0038 s         &  0.0073 s   & 0.0169 s    &   0.0463 s    & 0.1290 s & 0.4157 s         \\
		MFDE ($m=3$)     &     0.0049 s       &       0.0061 s     &  0.0097 s   &   0.0211 s  & 0.0541 s      &   0.1501 s   & 0.4945 s     \\
		MDE ($m=2$)     &    0.0028 s        &    0.0041 s        &  0.0078 s   &  0.0176 s   &    0.0478 s   &    0.1336 s   &   0.4189 s  \\
		MDE ($m=3$)          &  0.0053 s          &  0.0070 s          &   0.0111 s  & 0.0224 s    &   0.0598 s    &     0.1673 s&   0.5446 s     \\
		MSE ($m=2$)     &undefined at all scales&undefined at several scales& 0.0113 s       & 0.0743 s      &    0.7031 s       &       6.0879 s &72.1888 s  \\
		MSE ($m=3$)     &undefined at all scales&undefined at all scales&undefined at several scales&  0.0681 s    &  0.6546 s            &   5.6362 s &62.3229 s     \\
		
	\end{tabular}
\end{table*}

\subsection{Neurological Datasets}
In the physiological complexity literature, it is hypothesized that healthy conditions correspond to more complex states due to their ability to adapt to adverse conditions, exhibiting long range correlations, and rich variability at multiple scales, while aged and diseased individuals demonstrate complexity loss. That is, they lose the capability to adapt to such adverse conditions \cite{costa2005multiscale}. Therefore, we employ MFDE, compared with MDE and MSE, to characterize different pathological states using several neurological datasets. Note that we use these standard datasets only to evaluate the complexity methods, not to compete with other signal processing approaches.
	
\begin{figure*}
	\centering
	\includegraphics[width=18cm]{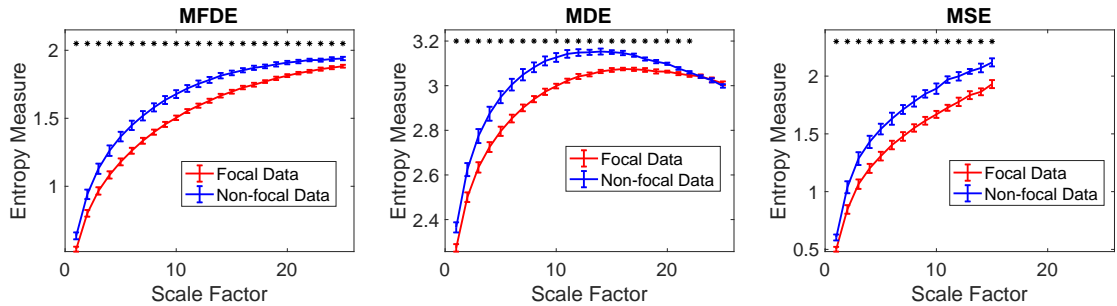}
	\caption{Mean value and SD of the results obtained by the MFDE, MDE, and MSE computed from the focal and non-focal EEGs. The scale factors with \textit{p}-values between 0.01 and 0.05, and smaller than 0.01 are respectively shown with + and *. The MSE values are undefined at high scale factors. The MSE values are undefined at high scale factors.}
	\label{N_63_EEG_Ralph}
\end{figure*}

\textit{1) Dataset of Focal and Non-focal Electroencephalograms (EEGs)}: The ability of the MFDE, MDE, and MSE techniques to distinguish the focal from non-focal signals is evaluated here. The results, depicted in Fig.~\ref{N_63_EEG_Ralph}, show that the non-focal signals are more complex than the focal ones. This fact is in agreement with previous studies \cite{andrzejak2012nonrandomness,sharma2015application}. Note that because the entropy-based methods are used for stationary signals \cite{richman2000physiological,azami2018amplitude}, we separated each signal into segments of length 2 s (1024 sample points) and applied the algorithms to each of them. The results demonstrate that all the techniques lead to the similar findings, albeit MDE and MFDE are significantly faster than MSE ones, as illustrated in Subsection \ref{SynthSiG}. It should be mentioned that the average entropy values over 2 channels for these bivariate EEG signals are reported for these univariate complexity techniques.  

The non-parametric Mann-Whitney \textit{U}-test was employed to evaluate the differences between results for focal vs.~non-focal signals at each scale factor. In this study, the scale factors with \textit{p}-values between 0.01 and 0.05, and smaller than 0.01 are respectively shown with + and *. The \textit{p}-values demonstrate that MFDE is the only complexity method with significant differences at all scale factors, showing its advantage over MSE and MDE.

\begin{figure*}
	\centering
	\includegraphics[width=18cm]{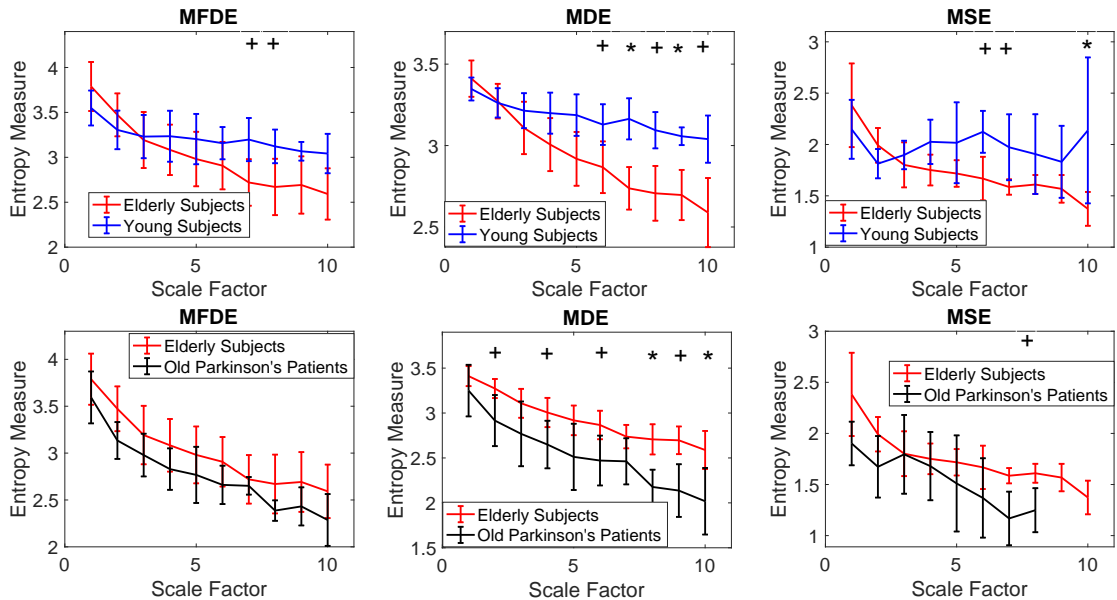}
	\caption{Mean value and SD of the results obtained by the MFDE, MDE, and MSE techniques computed from the young, elderly, and old Parkinson's subjects' stride interval recordings. The scale factors with \textit{p}-values between 0.01 and 0.05, and smaller than 0.01 are respectively shown with + and *. The MSE values are undefined at high scale factors.}
	\label{PD_Aging_Gait}
\end{figure*}

\textit{2) Dataset of Walking Stride Interval Time Series for Young, Elderly, and Parkinson's Disease (PD) Subjects}: As shown in Fig.~\ref{PD_Aging_Gait}, for most scale factors the average MFDE, MDE, and MSE values are smaller in elderly subjects compared with young subjects. This is consistent with those obtained by transfer entropy \cite{nemati2013respiration} and the fact that recordings from healthy young subjects correspond to more complex states due to their ability to adapt to adverse conditions, whereas older individuals' signals demonstrate complexity loss \cite{costa2005multiscale,fogedby1992phase,ahmed2011multivariate}. The results also show that the PD patients' stride interval recordings are less complex than those for the elderly subjects, which is in agreement with the fact that some diseases lead to lower complexity values \cite{costa2005multiscale,costa2002multiscale}. Since the length of each stride interval signal was between 200 to 700 samples, we did not separate the signals into smaller epochs.

The non-parametric Mann-Whitney \textit{U}-test was employed to evaluate the differences between results for young vs.~elderly individuals and elderly vs.~PD patients at each scale factor. The \textit{p}-values demonstrate that the best algorithm for the discrimination of PD from elderly subjects and elderly from young persons is MDE.

\begin{figure*}
	\centering
	\includegraphics[width=18cm]{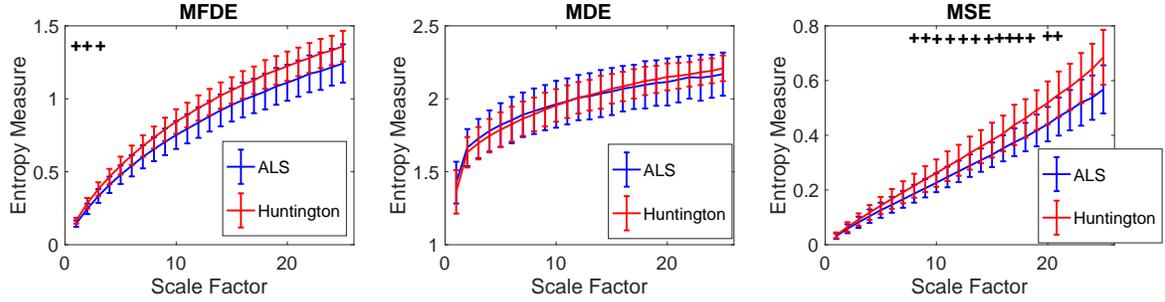}
	\caption{Mean value and SD of results obtained by the MFDE, MDE, and MSE techniques computed from the HD and ALS subjects' stride interval recordings. The scale factors with \textit{p}-values between 0.01 and 0.05, and smaller than 0.01 are respectively shown with + and *. }
	\label{HD_ALS_Gait}
\end{figure*}

\textit{3) Dataset of Walking Stride Interval Time Series for Huntington's Disease (HD) vs.~Amyotrophic Lateral Sclerosis (ALS) Patients}: Due to their long length, the signals were separated into epochs with length 3 s. The MFDE- and MSE-based results, depicted in Fig.~\ref{HD_ALS_Gait}, show that the stride interval fluctuations for HD are more complex than those for the ALS patients walking without any wheelchair or assistive device for mobility. This is in agreement with \cite{hausdorff2000dynamic,hausdorff1996fractal}. The \textit{p}-values show that both MFDE and MSE, unlike MDE, significantly discriminated the ALS from HD patients.

\begin{figure*}
	\centering
	\includegraphics[width=18cm]{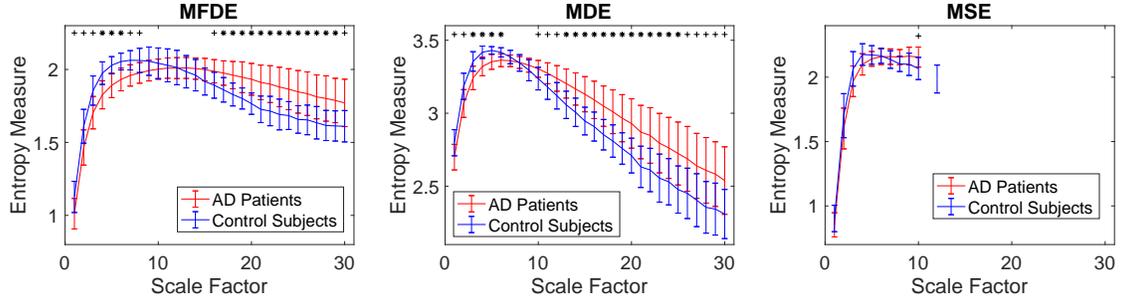}
	\caption{Mean value and SD of results of the MFDE, MDE, and MSE for 11 AD subjects vs.~11 age-matched controls. The scale factors with \textit{p}-values between 0.01 and 0.05, and smaller than 0.01 are respectively shown with + and *. The MSE values are undefined at high scale factors.}
	\label{EEG_AD}
\end{figure*}

\textit{4) Surface Electroencephalogram (EEG) Dataset in Alzheimer's Disease (AD)}: As the length of each EEG is 5 s, we do not separate the signals into smaller epochs. MFDE, MDE, and MSE were used to characterize the time series recorded from 11 AD patients vs.~11 age-matched healthy controls. The results are depicted in Fig.~\ref{EEG_AD}. The average of MFDE, MDE, and MSE values for AD patients was smaller than those for healthy controls at short-time scale factors, while the AD subjects' EEGs had larger entropy values at long-time scale factors. Herein, short-time (or low) scale factors mean the temporal scales that are smaller than or equal to the scale of crossing point of the curves for AD patients vs.~controls. Long-time (or high) scale factors denote the temporal scales that are larger than the scale of crossing point of the curves for AD patients vs.~controls. For example, short-time and long-time scale factors are 1-12 and 13-30, respectively, for MFE in Fig.~\ref{EEG_AD}. All the results are consistent with \cite{yang2013cognitive,labate2013entropic,escudero2015multiscale,azami2017univariate,azami2017refined}. Nevertheless, for MSE, unlike MDE and MFDE, values at high scale factors are undefined, showing an advantage of MFDE and MDE over MSE. Another advantage of MFDE and MDE over MSE is that these methods lead to larger differences at a number of temporal scale factors. Of note is that the average of the entropy values for all the channels is reported for the univariate multiscale entropy methods herein. 

\begin{figure*}
	\centering
	\includegraphics[width=18cm]{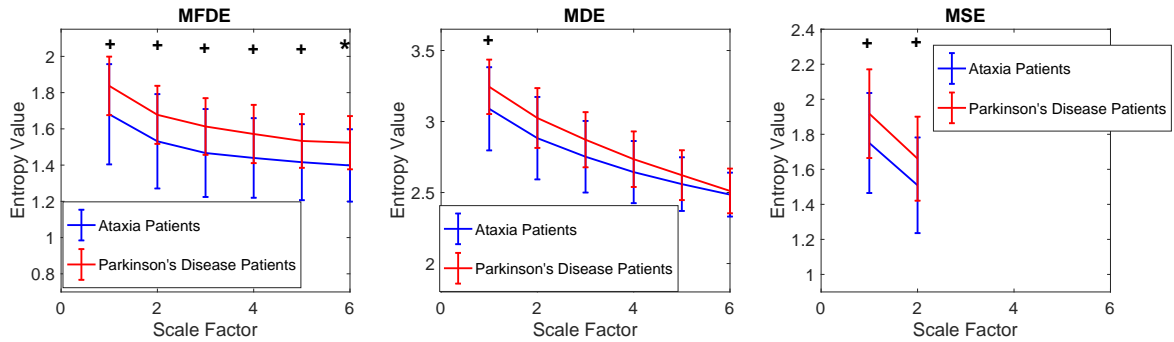}
	\caption{Mean value and SD of results obtained by the MFDE, MDE, and MSE techniques computed from the ataxia' and parkinsonism subjects' eye movement recordings. The scale factors with \textit{p}-values between 0.01 and 0.05, and smaller than 0.01 are respectively shown with + and *. The MSE values are undefined at high scale factors.}
	\label{Ataxia_PD_Eye_Movement}
\end{figure*}

\textit{5) Eye Movement Dataset for Parkinsonism vs.~Ataxia Patients}: To deal with the stationarity of signals, we separated each signal into epochs with length 1 s. The mean and SD of MFDE, MDE, and MSE values for parkinsonism vs.~ataxia patients are depicted in Fig.~\ref{Ataxia_PD_Eye_Movement}. The results show that the mean values for all the complexity methods computed from the parkinsonism subjects are higher than those recorded from the ataxia patients. This is consistent with the fact that oculomotor impairment is dramatic and a core clinical feature of cerebellar ataxia, whereas eye movement abnormalities in Parkinson's disease are relatively mild.

Like the other results, the MSE values are undefined at high scale factors. The Mann-Whitney \textit{U}-test \textit{p}-values show that only MFDE was significantly different in parkinsonism and ataxia patients across the range of scale factors. This shows that where the mean value of a time series noticeably changes along the signal, MFDE may be better than MSE and MDE in detecting different states of physiological data.

On the whole, the results support that, in general, MDE and MFDE perform better than MSE. MSE achieves better \textit{p}-values for the discrimination of ALS vs.~HD subjects (Fig.~11), but in the other datasets, it fails because it cannot be computed. We also showed that MSE is considerably slower than MDE and MFDE in Table I. Thus, we recommend MFDE and MDE over MSE for the analysis of physiological recordings. Between MDE and MFDE, based on the \textit{p}-values, MDE was better than MFDE only for the dataset of walking stride interval signals for young, elderly, and PD subjects (Fig.~10). However, MFDE outperformed MDE for the characterization of three neurological datasets: 1) focal vs.~non-focal EEGs (Fig.~9); 2) stride interval fluctuations for Huntington's disease vs.~amyotrophic lateral sclerosis (Fig.~11); and 3) eye movement data for parkinsonism vs.~ataxia (Fig.~13). In addition, MFDE results for pink noise were more stable than those for MDE (Fig.~7). Furthermore, MFDE was slightly faster than MDE (Table I). In sum, the results indicate that MFDE was the fastest and most consistent technique to distinguish various dynamics of the synthetic and real data, especially when dealing with the presence of baseline wanders, or trends, in signals.

\section{Future Work}
In spite of the promising findings based on MFDE and MDE, these novel signal processing approaches should be employed on various physiological datasets with a higher number of subjects in order to evaluate their ability for detection of dynamical variability of different kinds of timer series.

The physiological nature of the findings for AD vs.~controls needs to be further investigated to understand why AD patients' EEGs are less complex at low scale factors while the controls' recording are less complex at high temporal scales. With regard to eye movement, the higher complexity signal in PD compared with ataxia can be coarsely explained by the fact that eye movements are more impaired in ataxia. However, in future work we hope to better understand more precisely how and why abnormalities seen in ataxia result in a lower complexity signal.

In this article, the most commonly used coarse-graining process was used \cite{costa2005multiscale,azami2017refined_MFE,azami2017refined,morabito2012multivariate}. The alternative coarse-graining processes based on empirical mode decomposition and finite impulse response (FIR) filters \cite{azami2018coarse} can be employed instead of the classical implementation of coarse-graining process used herein. Refined composite MFDE based on refined composite MDE \cite{azami2017refined} can be proposed for very short univariate signals. The multivariate extension of MFDE dealing with both the time and spatial domain at the same time can also be developed.

\section{Conclusions}
In this paper, we introduced MFDE to quantify the complexity of time series based on their fluctuation-based dispersion patterns. The results on synthetic data showed that MFDE, MDE, and MSE lead to similar findings although MSE values were undefined at high scales. This fact, together with their much faster computation time, makes us recommend MFDE and MDE over MSE for the analysis of biomedical signals. Based on the Mann-Whitney \textit{U}-test \textit{p}-values, MDE outperformed MFDE only for the dataset of walking stride interval signals for young, elderly, and PD subjects. Both the MDE and MFDE methods significantly discriminated the AD patients from healthy controls. However, MFDE was better than MDE for the characterization of three neurological datasets: 1) focal vs.~non-focal EEGs; 2) stride interval fluctuations for Huntington's disease vs.~amyotrophic lateral sclerosis; and 3) eye movement data for Parkinson's disease vs.~ataxia, potentially because MFDE is robust to changes in the mean value of a time series, as seen in the eye movement dataset. Additionally, MFDE, compared with MDE, led to more stable entropy values over the scale factors for pink noise. These observations suggest that MFDE may be better than MSE and MDE in detecting different states of synthetic and physiological recordings. We expect MFDE, in addition to MDE, to be widely used for the characterization of different physiologic data in various neurological diseases.

\section*{Acknowledgment}
We would like to thank Dr. Pedro Espino (Hospital Clinico San Carlos, Madrid, Spain) for his help in the recording and selection of EEG epochs. We would like to thank Dr. Jeremy Schmahmann, Dr. Albert Hung, and Dr. Christopher Stephen for their help in recruiting research participants, and Mary Donovan for her help in collecting eye movement data from participants at MGH. We also thank Dr. Daniel Ab{\'a}solo (University of Surrey, Guildford, UK) for making the Alzheimer's disease dataset available. G. Sapiro and Z. Chang are partially supported by NSF, NIH, and DoD.


\begin{thebibliography}{10}
	
	\bibitem{pincus1991approximate}
	S.~M. Pincus, ``Approximate entropy as a measure of system complexity.,'' {\em
		Proceedings of the National Academy of Sciences}, vol.~88, no.~6,
	pp.~2297--2301, 1991.
	
	\bibitem{richman2000physiological}
	J.~S. Richman and J.~R. Moorman, ``Physiological time-series analysis using
	approximate entropy and sample entropy,'' {\em American Journal of
		Physiology-Heart and Circulatory Physiology}, vol.~278, no.~6,
	pp.~H2039--H2049, 2000.
	
	\bibitem{bandt2002permutation}
	C.~Bandt and B.~Pompe, ``Permutation entropy: a natural complexity measure for
	time series,'' {\em Physical review letters}, vol.~88, no.~17, p.~174102,
	2002.
	
	\bibitem{rostaghi2016dispersion}
	M.~Rostaghi and H.~Azami, ``Dispersion entropy: A measure for time series
	analysis,'' {\em IEEE Signal Processing Letters}, vol.~23, no.~5,
	pp.~610--614, 2016.
	
	\bibitem{shannon2001mathematical}
	C.~E. Shannon, ``A mathematical theory of communication,'' {\em ACM SIGMOBILE
		Mobile Computing and Communications Review}, vol.~5, no.~1, pp.~3--55, 2001.
	
	\bibitem{sanei2007eeg}
	S.~Sanei and J.~A. Chambers, {\em {EEG} signal processing}.
	\newblock John Wiley \& Sons, 2007.
	
	\bibitem{fogedby1992phase}
	H.~C. Fogedby, ``On the phase space approach to complexity,'' {\em Journal of
		statistical physics}, vol.~69, no.~1-2, pp.~411--425, 1992.
	
	\bibitem{costa2005multiscale}
	M.~Costa, A.~L. Goldberger, and C.-K. Peng, ``Multiscale entropy analysis of
	biological signals,'' {\em Physical Review E}, vol.~71, no.~2, p.~021906,
	2005.
	
	\bibitem{zhang1991complexity}
	Y.-C. Zhang, ``Complexity and 1/f noise. a phase space approach,'' {\em Journal
		de Physique I}, vol.~1, no.~7, pp.~971--977, 1991.
	
	\bibitem{azami2017refined_MFE}
	H.~Azami, A.~Fern{\'a}ndez, and J.~Escudero, ``Refined multiscale fuzzy entropy
	based on standard deviation for biomedical signal analysis,'' {\em Medical \&
		Biological Engineering \& Computing}, vol.~55, no.~11, pp.~2037--2052, 2017.
	
	\bibitem{costa2002multiscale}
	M.~Costa, A.~L. Goldberger, and C.-K. Peng, ``Multiscale entropy analysis of
	complex physiologic time series,'' {\em Physical Review Letters}, vol.~89,
	no.~6, p.~068102, 2002.
	
	\bibitem{humeau2015multiscale}
	A.~Humeau-Heurtier, ``The multiscale entropy algorithm and its variants: A
	review,'' {\em Entropy}, vol.~17, no.~5, pp.~3110--3123, 2015.
	
	\bibitem{bar1997dynamics}
	Y.~Bar-Yam, {\em Dynamics of complex systems}, vol.~213.
	\newblock Addison-Wesley Reading, MA, 1997.
	
	\bibitem{silva2015multiscale}
	L.~E.~V. Silva, B.~C.~T. Cabella, U.~P. da~Costa~Neves, and L.~O.~M. Junior,
	``Multiscale entropy-based methods for heart rate variability complexity
	analysis,'' {\em Physica A: Statistical Mechanics and its Applications},
	vol.~422, pp.~143--152, 2015.
	
	\bibitem{morabito2012multivariate}
	F.~C. Morabito, D.~Labate, F.~La~Foresta, A.~Bramanti, G.~Morabito, and
	I.~Palamara, ``Multivariate multi-scale permutation entropy for complexity
	analysis of {Alzheimer's} disease {EEG},'' {\em Entropy}, vol.~14, no.~7,
	pp.~1186--1202, 2012.
	
	\bibitem{wu2016refined}
	S.-D. Wu, C.-W. Wu, and A.~Humeau-Heurtier, ``Refined scale-dependent
	permutation entropy to analyze systems complexity,'' {\em Physica A:
		Statistical Mechanics and its Applications}, vol.~450, pp.~454--461, 2016.
	
	\bibitem{azami2018amplitude}
	H.~Azami and J.~Escudero, ``Amplitude-and fluctuation-based dispersion
	entropy,'' {\em Entropy}, vol.~20, no.~3, p.~210, 2018.
	
	\bibitem{azami2017refined}
	H.~Azami, M.~Rostaghi, D.~Ab{\'a}solo, and J.~Escudero, ``Refined composite
	multiscale dispersion entropy and its application to biomedical signals,''
	{\em IEEE Transactions on Biomedical Engineering}, vol.~64, no.~12,
	pp.~2872--2879, 2017.
	
	\bibitem{azami2018coarse}
	H.~Azami and J.~Escudero, ``Coarse-graining approaches in univariate multiscale
	sample and dispersion entropy,'' {\em Entropy}, vol.~20, no.~2, p.~138, 2018.
	
	\bibitem{goldberger2014complexity}
	A.~L. Goldberger and M.~D. Costa, ``Complexity based methods and systems for
	detecting depression,'' Aug.~7 2014.
	\newblock US Patent App. 14/233,999.
	
	\bibitem{chung2013multiscale}
	C.-C. Chung, J.-H. Kang, R.-Y. Yuan, D.~Wu, C.-C. Chen, N.-F. Chi, P.-C. Chen,
	and C.-J. Hu, ``Multiscale entropy analysis of electroencephalography during
	sleep in patients with parkinson disease,'' {\em Clinical EEG and
		neuroscience}, vol.~44, no.~3, pp.~221--226, 2013.
	
	\bibitem{rahman2018sleep}
	M.~M. Rahman, M.~I.~H. Bhuiyan, and A.~R. Hassan, ``Sleep stage classification
	using single-channel eog,'' {\em Computers in biology and medicine},
	vol.~102, pp.~211--220, 2018.
	
	\bibitem{miskovic2019changes}
	V.~Miskovic, K.~J. MacDonald, L.~J. Rhodes, and K.~A. Cote, ``Changes in eeg
	multiscale entropy and power-law frequency scaling during the human sleep
	cycle,'' {\em Human brain mapping}, vol.~40, no.~2, pp.~538--551, 2019.
	
	\bibitem{azami2017multiscale}
	H.~Azami, E.~Kinney-lang, A.~Ebied, A.~Fern{\'a}ndez, and J.~Escudero,
	``Multiscale dispersion entropy for the regional analysis of resting-state
	magnetoencephalogram complexity in alzheimer's disease,'' in {\em Engineering
		in Medicine and Biology Society (EMBC), 2017 39th Annual International
		Conference of the IEEE}, pp.~3182--3185, IEEE, 2017.
	
	\bibitem{hu2001effect}
	K.~Hu, P.~C. Ivanov, Z.~Chen, P.~Carpena, and H.~E. Stanley, ``Effect of trends
	on detrended fluctuation analysis,'' {\em Physical Review E}, vol.~64, no.~1,
	p.~011114, 2001.
	
	\bibitem{wu2007trend}
	Z.~Wu, N.~E. Huang, S.~R. Long, and C.-K. Peng, ``On the trend, detrending, and
	variability of nonlinear and nonstationary time series,'' {\em Proceedings of
		the National Academy of Sciences}, vol.~104, no.~38, pp.~14889--14894, 2007.
	
	\bibitem{peng1995quantification}
	C.-K. Peng, S.~Havlin, H.~E. Stanley, and A.~L. Goldberger, ``Quantification of
	scaling exponents and crossover phenomena in nonstationary heartbeat time
	series,'' {\em Chaos: An Interdisciplinary Journal of Nonlinear Science},
	vol.~5, no.~1, pp.~82--87, 1995.
	
	\bibitem{kaffashi2008effect}
	F.~Kaffashi, R.~Foglyano, C.~G. Wilson, and K.~A. Loparo, ``The effect of time
	delay on approximate \& sample entropy calculations,'' {\em Physica D:
		Nonlinear Phenomena}, vol.~237, no.~23, pp.~3069--3074, 2008.
	
	\bibitem{humeau2016refined}
	A.~Humeau-Heurtier, C.-W. Wu, S.~De~Wu, M.~Guillaume, and P.~Abraham, ``Refined
	multiscale hilbert-huang spectral entropy and its application to central and
	peripheral cardiovascular data,'' {\em IEEE Transactions on Biomedical
		Engineering}, pp.~1--11, 2016.
	
	\bibitem{azami2016improved}
	H.~Azami and J.~Escudero, ``Improved multiscale permutation entropy for
	biomedical signal analysis: Interpretation and application to
	electroencephalogram recordings,'' {\em Biomedical Signal Processing and
		Control}, vol.~23, pp.~28--41, 2016.
	
	\bibitem{lam2011preserving}
	J.~Lam, ``Preserving useful info while reducing noise of physiological signals
	by using wavelet analysis,'' pp.~1--20, 2011.
	
	\bibitem{houdrehigh2016}
	C.~Houdr{\'e}, D.~M. Mason, P.~Reynaud-Bouret, and J.~Rosinski, {\em High
		Dimensional Probability VII}.
	\newblock Springer, 2016.
	
	\bibitem{labate2013entropic}
	D.~Labate, F.~La~Foresta, G.~Morabito, I.~Palamara, and F.~C. Morabito,
	``Entropic measures of {EEG} complexity in {Alzheimer's} disease through a
	multivariate multiscale approach,'' {\em Sensors Journal, IEEE}, vol.~13,
	no.~9, pp.~3284--3292, 2013.
	
	\bibitem{yang2013cognitive}
	A.~C. Yang, S.-J. Wang, K.-L. Lai, C.-F. Tsai, C.-H. Yang, J.-P. Hwang, M.-T.
	Lo, N.~E. Huang, C.-K. Peng, and J.-L. Fuh, ``Cognitive and neuropsychiatric
	correlates of {EEG} dynamic complexity in patients with {Alzheimer's}
	disease,'' {\em Progress in Neuro-Psychopharmacology and Biological
		Psychiatry}, vol.~47, pp.~52--61, 2013.
	
	\bibitem{andrzejak2012nonrandomness}
	R.~G. Andrzejak, K.~Schindler, and C.~Rummel, ``Nonrandomness, nonlinear
	dependence, and nonstationarity of electroencephalographic recordings from
	epilepsy patients,'' {\em Physical Review E}, vol.~86, no.~4, p.~046206,
	2012.
	
	\bibitem{hausdorff1996fractal}
	J.~M. Hausdorff, P.~L. Purdon, C.~Peng, Z.~Ladin, J.~Y. Wei, and A.~L.
	Goldberger, ``Fractal dynamics of human gait: stability of long-range
	correlations in stride interval fluctuations,'' {\em Journal of Applied
		Physiology}, vol.~80, no.~5, pp.~1448--1457, 1996.
	
	\bibitem{acharya2018characterization}
	U.~R. Acharya, Y.~Hagiwara, S.~N. Deshpande, S.~Suren, J.~E.~W. Koh, S.~L. Oh,
	N.~Arunkumar, E.~J. Ciaccio, and C.~M. Lim, ``Characterization of focal eeg
	signals: a review,'' {\em Future Generation Computer Systems}, 2018.
	
	\bibitem{Ralph_Data}
	{http://ntsa.upf.edu/}
	
	\bibitem{hausdorff2000dynamic}
	J.~M. Hausdorff, A.~Lertratanakul, M.~E. Cudkowicz, A.~L. Peterson, D.~Kaliton,
	and A.~L. Goldberger, ``Dynamic markers of altered gait rhythm in amyotrophic
	lateral sclerosis,'' {\em Journal of applied physiology}, vol.~88, no.~6,
	pp.~2045--2053, 2000.
	
	\bibitem{XYZ}
	{https://www.physionet.org/physiobank/database/gaitdb}
	
	\bibitem{goldfarb1984gait}
	B.~Goldfarb and S.~Simon, ``Gait patterns in patients with amyotrophic lateral
	sclerosis.,'' {\em Archives of physical medicine and rehabilitation},
	vol.~65, no.~2, pp.~61--65, 1984.
	
	\bibitem{ALS_HD}
	{https://physionet.org/physiobank/database/gaitndd/}
	
	\bibitem{alzheimer20172017}
	{Alzheimer's Association}, ``2017 {Alzheimer's} disease facts and figures,''
	{\em Alzheimer's \& Dementia}, vol.~13, no.~4, pp.~325--373, 2017.
	
	\bibitem{wilson2012natural}
	R.~S. Wilson, E.~Segawa, P.~A. Boyle, S.~E. Anagnos, L.~P. Hizel, and D.~A.
	Bennett, ``The natural history of cognitive decline in {Alzheimer's}
	disease,'' {\em Psychology and Aging}, vol.~27, no.~4, p.~1008, 2012.
	
	\bibitem{dauwels2010diagnosis}
	J.~Dauwels, F.~Vialatte, and A.~Cichocki, ``Diagnosis of {Alzheimer's} disease
	from {EEG} signals: where are we standing?,'' {\em Current Alzheimer
		Research}, vol.~7, no.~6, pp.~487--505, 2010.
	
	\bibitem{bhat2015clinical}
	S.~Bhat, U.~R. Acharya, N.~Dadmehr, and H.~Adeli, ``Clinical neurophysiological
	and automated {EEG}-based diagnosis of the {Alzheimer's} disease,'' {\em
		European Neurology}, vol.~74, no.~3-4, pp.~202--210, 2015.
	
	\bibitem{abasolo2006entropy}
	D.~Ab{\'a}solo, R.~Hornero, P.~Espino, D.~Alvarez, and J.~Poza, ``Entropy
	analysis of the {EEG} background activity in {Alzheimer's} disease
	patients,'' {\em Physiological measurement}, vol.~27, no.~3, p.~241, 2006.
	
	\bibitem{abasolo2008approximate}
	D.~Ab{\'a}solo, J.~Escudero, R.~Hornero, C.~G{\'o}mez, and P.~Espino,
	``Approximate entropy and auto mutual information analysis of the
	electroencephalogram in {Alzheimer's} disease patients,'' {\em Medical \&
		biological engineering \& computing}, vol.~46, no.~10, pp.~1019--1028, 2008.
	
	\bibitem{escudero2006analysis}
	J.~Escudero, D.~Ab{\'a}solo, R.~Hornero, P.~Espino, and M.~L{\'o}pez,
	``Analysis of electroencephalograms in {Alzheimer's} disease patients with
	multiscale entropy,'' {\em Physiological measurement}, vol.~27, no.~11,
	p.~1091, 2006.
	
	\bibitem{tombaugh1992mini}
	T.~N. Tombaugh and N.~J. McIntyre, ``The mini-mental state examination: a
	comprehensive review,'' {\em Journal of the American Geriatrics Society},
	vol.~40, no.~9, pp.~922--935, 1992.
	
	\bibitem{buttner98}
	N.~Buttner, D.~Geschwind, J.~C. Jen, S.~Perlman, S.~M. Pulst, and R.~W. Baloh,
	``Oculomotor phenotypes in autosomal dominant ataxias,'' {\em Arch Neurol},
	vol.~55, pp.~1353--7, Oct 1998.
	
	\bibitem{vidailhet94}
	M.~Vidailhet, S.~Rivaud, N.~Gouider-Khouja, B.~Pillon, A.~M. Bonnet,
	B.~Gaymard, Y.~Agid, and C.~Pierrot-Deseilligny, ``Eye movements in
	parkinsonian syndromes,'' {\em Ann Neurol}, vol.~35, pp.~420--6, Apr 1994.
	
	\bibitem{rottach96}
	K.~G. Rottach, D.~E. Riley, A.~O. DiScenna, A.~Z. Zivotofsky, and R.~J. Leigh,
	``Dynamic properties of horizontal and vertical eye movements in parkinsonian
	syndromes,'' {\em Ann Neurol}, vol.~39, pp.~368--77, Mar 1996.
	
	\bibitem{VIDEO}
	{https://www.ri.cmu.edu/publications/intraface/}
	
	\bibitem{sharma2015application}
	R.~Sharma, R.~B. Pachori, and U.~R. Acharya, ``Application of entropy measures
	on intrinsic mode functions for the automated identification of focal
	electroencephalogram signals,'' {\em Entropy}, vol.~17, no.~2, pp.~669--691,
	2015.
	
	\bibitem{nemati2013respiration}
	S.~Nemati, B.~A. Edwards, J.~Lee, B.~Pittman-Polletta, J.~P. Butler, and
	A.~Malhotra, ``Respiration and heart rate complexity: effects of age and
	gender assessed by band-limited transfer entropy,'' {\em Respiratory
		physiology \& neurobiology}, vol.~189, no.~1, pp.~27--33, 2013.
	
	\bibitem{ahmed2011multivariate}
	M.~U. Ahmed and D.~P. Mandic, ``Multivariate multiscale entropy: A tool for
	complexity analysis of multichannel data,'' {\em Physical Review E}, vol.~84,
	no.~6, p.~061918, 2011.
	
	\bibitem{escudero2015multiscale}
	J.~Escudero, E.~Acar, A.~Fern{\'a}ndez, and R.~Bro, ``Multiscale entropy
	analysis of resting-state magnetoencephalogram with tensor factorisations in
	{Alzheimer's} disease,'' {\em Brain research bulletin}, vol.~119,
	pp.~136--144, 2015.
	
	\bibitem{azami2017univariate}
	H.~Azami, D.~Ab{\'a}solo, S.~Simons, and J.~Escudero, ``Univariate and
	multivariate generalized multiscale entropy to characterise {EEG} signals in
	{Alzheimer's} disease,'' {\em Entropy}, vol.~19, no.~1, p.~31, 2017.
	
\end{thebibliography}
\end{document}